\documentclass[twoside,10pt]{article}
\usepackage[dvipsnames]{xcolor}

\usepackage{natbib}
\usepackage{custom}  

\usepackage{pstricks, pst-plot}	
\usepackage{graphicx} 
\usepackage{wrapfig}  
\usepackage[figuresright]{rotating}

\usepackage{amsmath}
\usepackage{amssymb}

\usepackage[english]{babel}
\usepackage{blindtext}
\usepackage{algorithm, algorithmicx, algpseudocode}
\usepackage{caption}
\usepackage{subcaption}
\usepackage{braket}

\newcommand{\argmax}{\mathop{\mathrm{argmax}}}

\newcommand{\ubar}[1]{\text{\b{$#1$}}}
\newcommand{\matern}[0]{Mat\'ern\ }

\title{Bayesian Optimization Priors for Efficient Variational Quantum Algorithms}

 \author[a,b,c]{Farshud Sorourifar}
 \author[b,c]{Diana Chamaki }
 \author[b]{Norm M. Tubman}
 \author[a]{Joel A. Paulson}
 \author[b,c,d]{\\David E. Bernal Neira}

 \affil[a]{\RaggedRight Department of Chemical and Biomolecular Engineering, The Ohio State University, Columbus, OH, USA}
 \affil[b]{Quantum Artificial Intelligence Laboratory (QuAIL), NASA Ames Research Center, Moffett Field, CA, USA}
 \affil[c]{USRA Research Institute for Advanced Computer Science, Mountain View, CA, USA}
 \affil[d]{Davidson School of Chemical Engineering, West Lafayette, IN, USA}
 \email{dbernaln@purdue.edu}

 \HeaderTitle{Improved BO methods for VQAs}
 \HeaderAuthor{F. Sorourifar et al.}

\begin{document}

\maketitle             
\thispagestyle{empty}  


\begin{abstract} 
    Quantum computers currently rely on a hybrid quantum-classical approach known as Variational Quantum Algorithms (VQAs) to solve problems.  
    Still, there are several challenges with VQAs on the classical computing side:
    it corresponds to a black-box optimization problem that is generally non-convex, the observations from the quantum hardware are noisy, and the quantum computing time is expensive. The first point is inherent to the problem structure; as a result, it requires the classical part of VQAs to be solved using global optimization strategies. However, there is a trade-off between cost and accuracy; typically, quantum computers return a set of bit strings, where each bitstring is referred to as a shot. The probabilistic nature of quantum computing (QC) necessitates many shots to measure the circuit accurately. Since QC time is charged per shot, reducing the number of shots yields cheaper and less accurate observations. Recently, there has been increasing interest in using basic Bayesian optimization (BO) methods to globally optimize quantum circuit parameters. This work proposes two modifications to the basic BO framework to provide a shot-efficient optimization strategy for VQAs. Specifically, we provide a means to place a prior on the periodicity of the rotation angles and a framework to place a topological prior using few-shot quantum circuit observations. We demonstrate the effectiveness of our proposed approach through an ablation study, showing that using both proposed features statistically outperforms a standard BO implementation within VQAs for computational chemistry simulations.       
\end{abstract}



\section{Introduction}

Quantum computing (QC) has been the subject of growing interest in chemical engineering, owing to its potential to solve computationally challenging problems, \cite{Bernal22}. However, being a nascent technology, it has a limited number of processing units (qubits) and states that rapidly decohere.  
To circumvent the quick decoherence and limited number of qubits in current devices, Variational Quantum Algorithms (VQAs) have been proposed, wherein a classical machine selects parameters for a quantum circuit representing a problem of interest, and a quantum machine evaluates it, \cite{Cerezo20}.
This circuit encodes the evolution of prepared qubits through a series of parameterized operators or gates.
The final qubit system state should follow a distribution representing the solution to a computational problem.
Physically, the quantum gate parameters represent rotation angles, which modify the quantum state of the qubits system. 
The final states of the qubits are then measured by projecting them into a set of classical bits (bitstring), with each bitstring measurement referred to as a shot. 
Thus, VQAs entail a feedback loop where a classical optimization algorithm selects the parameters for the quantum circuit based on a measure of the bitstrings to find the optimal parameters for a quantum circuit using the variational principle \cite{Pellow-Jarman23}.

Bayesian Optimization (BO) is a family of sample-efficient zeroth-order optimizers and has successfully solved various black-box problems, including VQAs.
BO's sample efficiency results from using observations from the quantum circuit to construct a statistical surrogate model known as a Gaussian Process (GP), which generalizes a multivariate normal distribution to function space (i.e., a probability distribution over an infinite set of functions that fit the data).
The GP's ability to quantify the model uncertainty allows us to systematically trade off between exploring the parameter space (e.g., learning how the parameters affect the objective) and exploiting promising regions (e.g., trying to improve on the best-observed parameters, often by sampling near the incumbent). 
BO has been gaining increased research interest for solving VQAs, spanning introductions to BO for VQAs, \cite{tibaldi23}, benchmarking, \cite{Ciavarella22}, and initialization strategies, \cite{Muller22,Tamiya2022}.
The authors in \cite{iannelli2021noisy} provide evidence that standard BO algorithms may benefit from lower-shot queries.

This work provides a background on the BO algorithm in \S \ref{sec:BO_prelims}.
In \S~\ref{sec:specialized_modifications}, we propose two principled modifications to the vanilla BO algorithm (i.e., a standard GP built from a \matern kernel with only a zero mean prior) to improve its efficiency in solving VQAs. Specifically, we propose a means to encoding a prior on the parameter's 2$\pi$ periodicity into the GP kernel function and a strategy for encoding a sample-based topological prior learned by fitting a second GP to low-shot measurements.  
In \S~\ref{sec:results}, we show through a simulation-based ablation study that these modifications can significantly improve BO performance on VQAs and provide concluding remarks in \S~\ref{sec:conclusion}.

\section{Bayesian Optimization Preliminaries}\label{sec:BO_prelims}

First, let $J(\theta)$ be the true circuit value evaluated at a given vector of rotation angles $\theta$. When we query the circuit, we are restricted to measuring noisy observations,
\begin{align}
    y(\theta, s) = J(\theta)+ \epsilon(s),
\end{align}
 where $\epsilon(s) \sim \mathcal{N}(0,\sigma^2(s))$, is an independent and identically distributed Gaussian noise, whose variance is dictated by the number of shots $s$ used to evaluate the circuit. We let $\epsilon =\epsilon(\bar{s})$ be the noise evaluated at the largest number of shots $\bar{s}$, which we are interested in using to observe the circuit. 
 BO begins with an initial data set $\mathcal{D}_0 = \{\theta_i,y_i\}_{i=1}^{\mathcal{I}}$ consisting of $\mathcal{I}$ initial observations, which can be used to build a statistical surrogate model. The choice of model in this framework is general, with the only requirement being that the model is \emph{statistical}, in the sense that it can quantify epistemic prediction uncertainty in terms of a covariance function.       
 This covariance function, coupled with the model's mean, is used to guide us in selecting sample points that balance exploration with exploitation through constructing and optimizing an acquisition function. Typically, this is done in between circuit observations and can be cheaply optimized using higher-order gradient methods. While many statistical models exist, the GPs are the most common choice due to their rigorous statistical quantification and non-parametric nature. In the following, we summarize the GP modeling and acquisition functions.  
 
\subsection{Gaussian Processes }
Here, we briefly introduce the GP and refer the interested reader to \cite{rasmussen2003gaussian} for a detailed treatment. We assume that the circuit has a GP prior 
of the form 
$f(\theta) \sim \mathcal{GP}(\mu_0, k)$ where 
$\mu_0: \vartheta \to \mathbb{R}$ is the prior mean and 
$k : \vartheta \times \vartheta \to \mathbb{R}$ is the prior covariance function. There are many possible choices for the covariance function, which we will discuss in the following section. Here, we formally define the \matern kernel function,
\begin{align}
k_{\nu}\left(\theta_i, \theta_j\right)=\frac{\sigma_f^2}{\Gamma(\nu) 2^{\nu-1}}\left(\frac{\sqrt{2 \nu}}{\ell} d\left(\theta_i, \theta_j\right)\right)^\nu K_\nu\left(\frac{\sqrt{2 \nu}}{\ell} d\left(\theta_i, \theta_j\right)\right),
\end{align}
where $\ell$ is a length-scale parameter, $\sigma_f^2$ i the measurement noise variance,   $\Gamma(\nu), K_\nu$ are the modified Bessel and gamma functions, and $d(\cdot)$ is a Euclidean distance function. The choice of $\nu$ is based on how smooth the function is believed to be, where larger values indicate a smoother function. Under the GP prior the $n$ function evaluations $y_{1:n}$ are jointly Gaussian with mean $[\mathbf{m}]_i = \mu_0(\theta_i)$, and covariance $[\mathbf{K}]_{i,j} = k(\theta_i,\theta_j)$ where $\mathbf{K} \in \mathbb{R}^{n \times n}$. This implies the corresponding function value $f(\theta)$ at any test point $\theta_{1:n}$ must be jointly Gaussian with $y_{1:n}$. Due to the properties of jointly Gaussian random variables, we find that the posterior distribution of the objective given all available noisy observations $p(f(\theta) | y_{1:n}, \theta_{1:n}, \theta)$, is Gaussian with the following mean and covariance
\begin{subequations} \label{eq:posterior-mean-variance}
\begin{align}
\mu_{n}(\theta) &= \mu_0(\theta) + \mathbf{k}(\theta)^\top (\mathbf{K} + \sigma^2 I_n ) ( y^f_{1:n} - \mathbf{m} ), \label{eq:mean}\\
\sigma^2_{n}(\theta) &= k(\theta,\theta) - \mathbf{k}(\theta)^\top (\mathbf{K} + \sigma^2 I_n)^{-1} \mathbf{k}(\theta) 
\label{eq:var}
\end{align}
\end{subequations}
where $[\mathbf{k}(\theta)]_i = k(\theta,\theta_i) $. With the mean and variance functions defined, we introduce the acquisition function.

\subsection{Acquisition Functions}
We can select a point $\theta$ that provides the most value for the circuit optimization by optimizing the acquisition function. For example, the lower confidence bound (LCB) acquisition function
\begin{equation}
\alpha_{LCB}(\theta) = \mu(\theta) - \sqrt{\beta}\sigma(\theta), 
\end{equation} 
balances exploration and exploitation by assigning an optimistic value to each candidate point. The exploitation term is represented by the mean and the exploration term by the standard deviation scaled by a parameter $\sqrt{\beta}$. Small values of $\beta$ result in an exploitative strategy, and large values of $\beta$ result in a more exploratory strategy. Other commonly used acquisition functions can be found in~\cite{frazier-tutorial}.

\section{Specialized Quantum Computing Priors}\label{sec:specialized_modifications}

\subsection{Periodic Parameter Prior}
While BO requires box constraints on the parameters, the periodic boundary conditions are not typically known or enforced. We can, however, use periodic kernels to codify the periodicity of the circuit measurements and uncertainty in the GP model using a periodic kernel \cite{periodicKernel}. The periodic kernel can be defined as
\begin{equation}
k_{\text {Periodic }}\left(\mathbf{\theta}, \mathbf{\theta}^{\prime}\right)=\sigma_f^2\exp \left(-2 \sum_i \frac{\sin ^2\left(\frac{\pi}{p}\left(\theta_i-\theta_i^{\prime}\right)\right)}{\ell_i}\right), 
\end{equation}
where $p$ is the period, and $\ell_i$ is the length-scale corresponding to the $i^{th}$ parameter. Typically, $p$ is treated as an unknown parameter that we fit to data since most optimization problems do not have known periodicity. Given that $p=2\pi$ for VQAs, we can fix this parameter during the model fitting. 

\begin{figure}[h]
    \centering
         \includegraphics[width=.45\textwidth,]{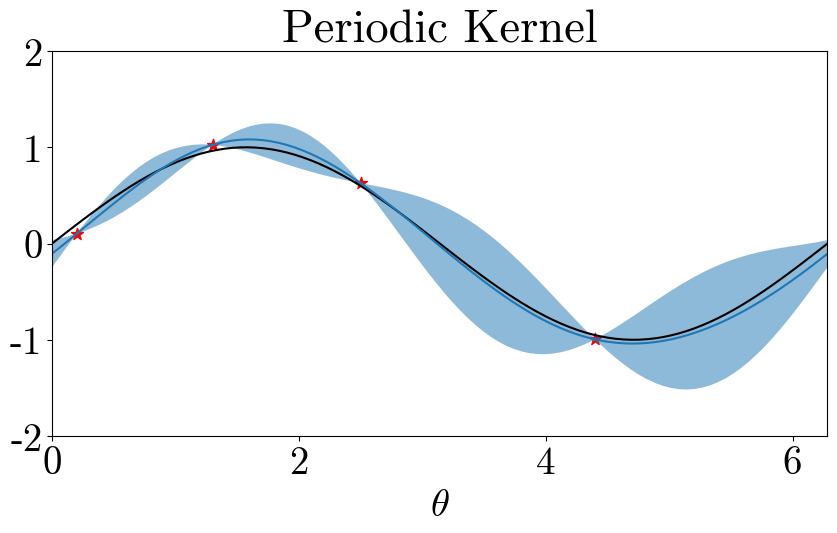}     
         \hspace{20pt}
        \includegraphics[width=.45\textwidth, ]{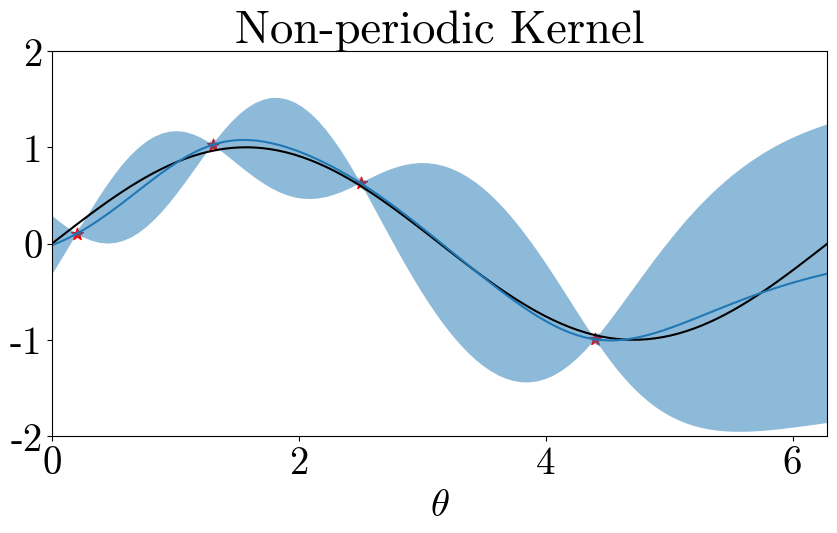}
     \caption{ Effect of a periodic (left) versus non-periodic (right) kernel on a periodic function. The periodic kernel has noticeably higher accuracy for $\theta>5$ due to recognizing the periodicity of the parameter space. 
     }
     \label{fig:ex_periodic}
\end{figure}

In Fig. \ref{fig:ex_periodic}, we provide an illustrative example of how knowledge of the periodic boundaries improves the surrogate model accuracy. The true function is represented as a solid black line, the data as red stars, and the mean and $95\%$ CI as a blue line and clouds, respectively. We bring attention to the region $\theta\in[4.7, 2\pi]$, where uncertainty is substantially lower for the periodic kernel due to a measurement near $\theta = 0.2$. 
The non-periodic GP would recommend the following sample near $\theta =2\pi$ to explore the high uncertainty region, thus wasting samples.          

\subsection{Topological Prior }
\begin{figure}[h]
    \centering
         \includegraphics[width=.45\textwidth]{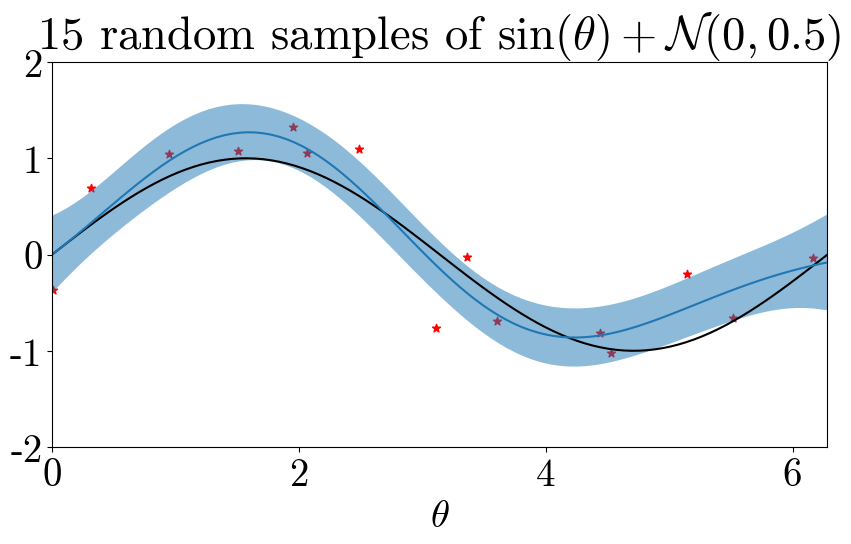}
         \hspace{20pt}
        \includegraphics[width=.45\textwidth]{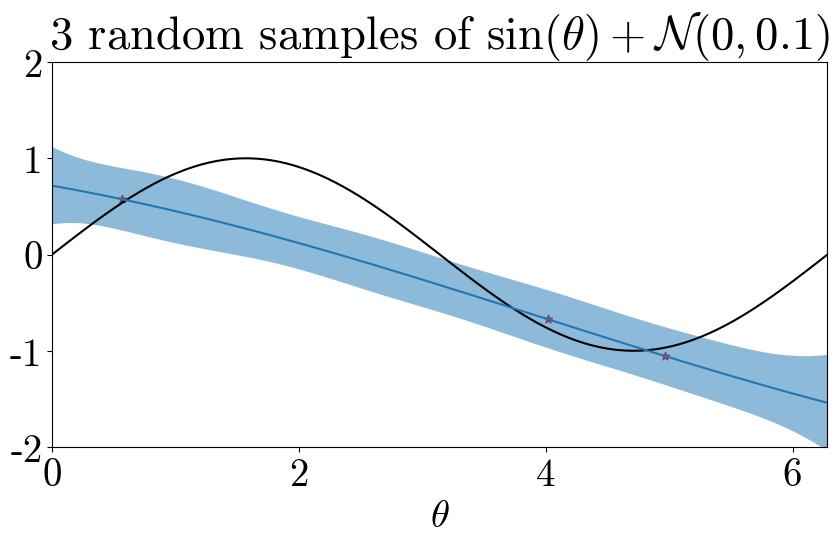}
     \caption{Effect of cheap and noisy (left) versus expensive and precise data (right). This illustrative example shows that samples taken with $\times$5 noise for $\times$1/5 cost may be more useful than fewer, more accurate samples.    
     }
     \label{fig:lsr_example}
\end{figure}

Although reducing the measured distribution of bitstrings to a single value, such as by taking the mean or conditional value at risk (CVaR), allows traditional optimization strategies to be easily adapted to solve VQAs, they also result in a loss of information. 
For example, consider the illustrative comparison in \ref{fig:lsr_example}, where we show how a larger volume (five times as many data points) of noisier data (standard deviation of the additive noise term is five times greater) may still result in a better fitting surrogate model than one fit to a smaller, although more precise, set of data. This illustrative example can be likened to a quantum circuit, where a user can select the number of shots for each measurement given a fixed shot budget.        

To use the low-shot measurements as a topological prior, we propose using a low-shot residual (LSR) inspired by the reference models in \cite{lu2021}, which we adapt for VQAs by exploiting the ability to regulate the number of shots per query. For notation simplicity, we now assume that $y(\theta,\bar{s}) = J(\theta)$, i.e., $\bar{s}$ is sufficiently high such that noise can be ignored. Given that we wish to minimize the output of a high-shot circuit $J(\theta)$ which uses $\bar{s}$ shots per circuit call, we assume that a fraction of the total shot budget $\gamma B$ may be spent on observations of a low-shot circuit $g(\theta)$, which uses $\ubar{s}$ shots per circuit call. We assume that the high-shot model can be defined as 
\begin{equation}\label{eq:true_lsr}
    J(\theta) = g(\theta)+\epsilon (\theta),
\end{equation}
where function $\epsilon (\theta)$ is a residual, or error, between the high and low-shot observations for a given $\theta$. However, this assumption must be satisfied by construction. Let $\mathcal{D}_0 =\{\theta_i, g_i\}^m$ be the set of $m$ low-shot observations obtained from spending $\gamma B$ uniformly over $\vartheta$, and $\mu_g(\theta), \sigma_g(\theta)$ be the mean and variance functions obtained from fitting a GP to $\mathcal{D}_0$. With the low-shot budget exhausted, $\mu_g(\theta)$ can be treated as a deterministic function that approximates $g(\theta)$, which is the topological prior used to improve learning $J(\theta)$. With each measurement $k$, we construct a data set $\mathcal{D}_k =\{\theta_i, J_i-\mu_g(\theta_i)\}^k$ to build a mean and variance function of the residual $\mu_{\epsilon}(\theta), \sigma_{\epsilon}(\theta)$, while satisfying \eqref{eq:true_lsr}. 

Through this construction, we can redefine equation \eqref{eq:mean} as
\begin{equation} \label{eq:lsr_mean}
    \mu_J(\theta) = \mu_g(\theta)+    \mu_\epsilon(\theta),
\end{equation}
and can optimize the circuit by minimizing $\mu_g(\theta)+\mu_\epsilon(\theta)$. 
Since the low-shot model can no longer reduce its variance, it serves no value in informing the exploration of the parameter space. Instead, we can formulate the LCB acquisition function for the LSR as 
\begin{equation}\label{eq:lsr_acq}
\alpha_{LCB, LSR}(\theta)=(\mu_g(\theta)+\mu_\epsilon(\theta))-\sqrt{\beta} \sigma_\epsilon(\theta),
\end{equation}
where we use the mean function $\mu_J = \mu_g(\theta)+\mu_\epsilon(\theta)$, but only use the residual standard deviation $\sigma_\epsilon$ for exploration. Note that the low-shot residual acquisition will need a larger exploration constant $\beta$, given that the residual magnitudes are significantly smaller than the means.      
We summarize the LSR-BO in Algorithm \ref{alg:lsr}. 

\begin{algorithm}[h]
\caption{Low-shot Residual Bayesian Optimization algorithm}
\begin{algorithmic}[1]
\State{\textbf{Initialize:} Number of shots $\bar{s},\ubar{s}$;  low-shot data $\mathcal{D}_0$ from spending $\gamma B$ on $g(\theta)$; total shot budget $B$; and GP priors $m_g, m_\epsilon$ and kernels $k_g, k_\epsilon$. Initialize the algorithm with $k=0, ~ B_k= \gamma B$}
\State{construct $\mu_g$ from $\mathcal{D}_0$}
\While{$B_k\leq B$}
\State{Solve \eqref{eq:lsr_acq} to find $\theta_{k+1}$ }
\State{Evaluate quantum circuit $J(\theta_{k+1})$}
\State{Construct GP surrogate model for $f(\theta)$ given available data $\mathcal{D}_{k+1}=\{ \theta_{i}, J(\theta_{i})-\mu_\epsilon(\theta_{i}) \}_{i=0}^{k+1}$}
\State{$B_{k+1}~\mathrel{+}=~\bar{s}$}
\EndWhile
\end{algorithmic}
\label{alg:lsr}
\end{algorithm}
\textbf{}

\section{Experimental Results}\label{sec:results}
\begin{figure}
\centering
\begin{minipage}{.46\textwidth}
  \centering
  \includegraphics[width=\linewidth]{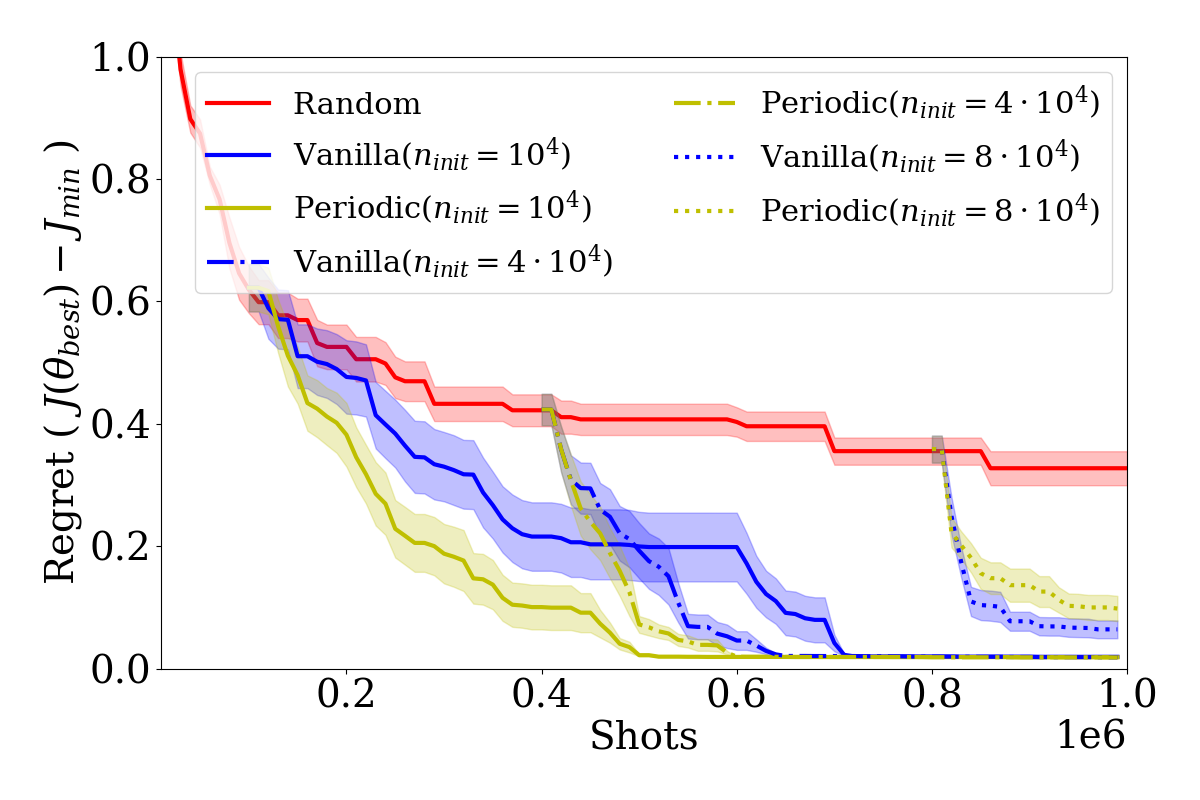}
  \captionof{figure}{Comparison of \matern versus periodic kernel-based GPs in BO, for multiple $\gamma$.  }
    \label{fig:kern_results}
\end{minipage}%
\hspace{20pt}
\begin{minipage}{.46\textwidth}
  \centering
  \includegraphics[width=\textwidth]{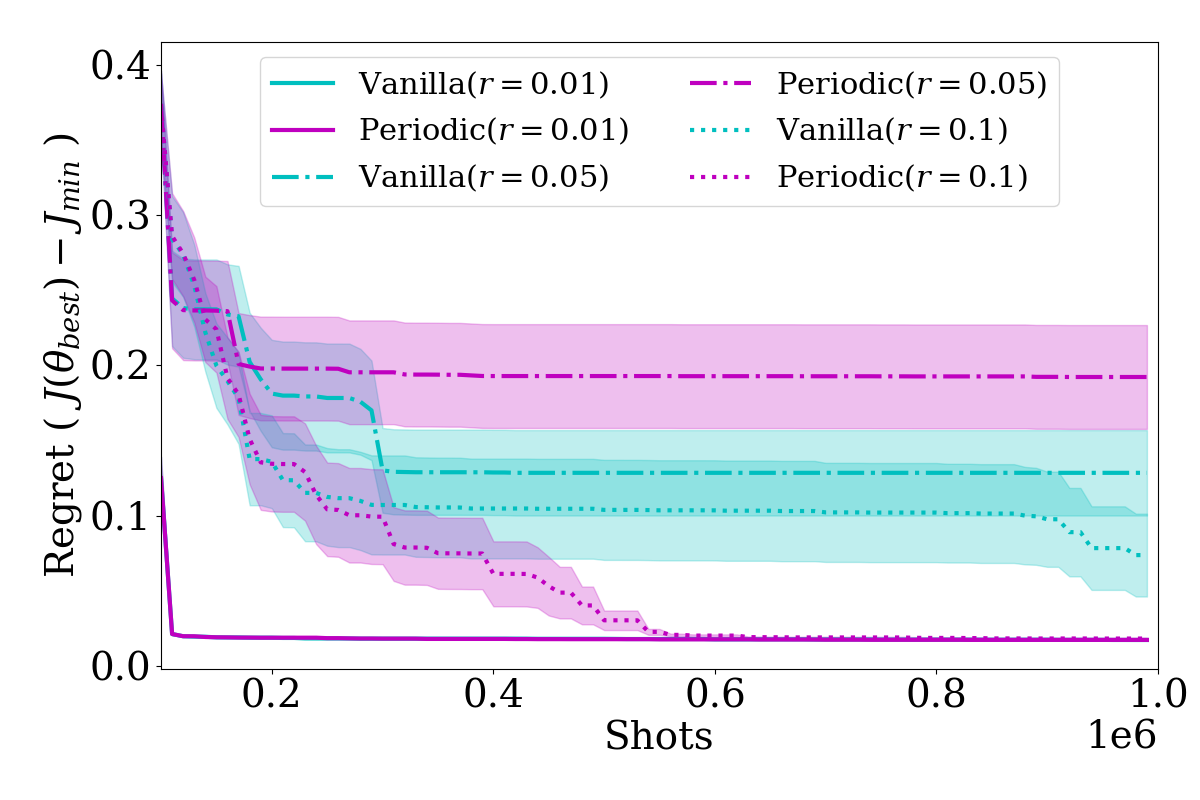}
  \captionof{figure}{Comparison of LSR-BO  using a \matern and periodic kernels, for different ratios of $r = \ubar{s}/\bar{s}$}
    \label{fig:lsr_experiment}
\end{minipage}
\end{figure}

In Fig.~\ref{fig:kern_results}, we compare the periodic kernel (yellow lines) versus a \matern kernel (blue lines) using statistical convergence on a hydrogen VQE problem using $\bar{s}=10000$ as the number of shots per an evaluation, and a total budget of $B=100\bar{s}$. Additionally, we vary the fraction of the total shot budget $\gamma=\{0.1, 0.4, 0.8, 1.0 \}$, where the initialization budget is spent on uniform random samples over the parameter space. Note that $\gamma = 1.0$ corresponds to a random sampling strategy since no budget is reserved for the BO algorithm. First, the random sampling strategy (shown in red) results in the worst performance, where smaller values of $\gamma$ improve convergence for the periodic kernel. In the \matern kernel, the trend isn't as apparent; convergence is faster with $\gamma=0.4$ (blue-dashed line) than $\gamma=0.1$ (blue-dotted line). We also note that as a larger budget is spent on random sampling, the performance difference between the kernels diminishes (and eventually reverses for $\gamma=0.8$) since the highly exploitative random sampling reduces the need for exploration, and thus, the benefits derived from the periodic kernel. Finally, since the random samples are highly exploitative, large $\gamma$ allows the Bayesian algorithms to focus on exploitation, evident by the faster convergence when switching from random to Bayesian sampling strategies.   

We present the results of the LSR-BO strategy in Fig. \ref{fig:lsr_experiment}, where we consider three values for the ratio of shots between the high and low shot circuits, $r = \ubar{s}/\bar{s} \in \{0.01, 0.05, 0.1\}$, where $\bar{s} = 10000$. From these results, we see a clear trend favoring residual models constructed from very few shots, where for small $r$ values, the algorithm finds near-optimal parameters on the high-shot circuit within the first few queries. However, larger $r$ values may diminish performance, as seen in the $r=0.1$ case with a vanilla kernel. While the results suggest that $\ubar{s}=1$ could provide the best performance, such a strategy would result in numerical issues. As shown in \eqref{eq:var}, the GP fitting requires a matrix inversion $(K-\sigma^2I_n)^{-1}$, which results in cubic scaling and increased risk of singularity as the number of data points grows.

\begin{figure}[h]
    \centering
         \includegraphics[width=.46\textwidth]{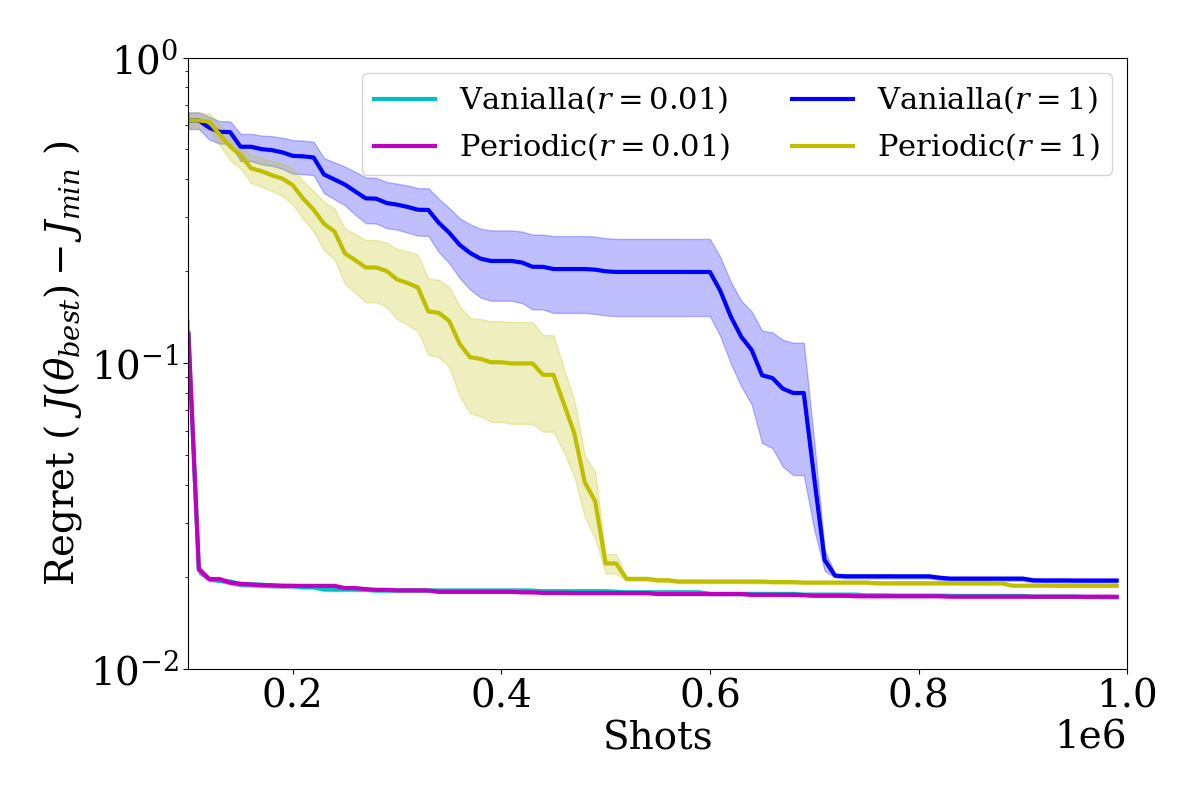}
        \hspace{20pt}
        \includegraphics[width=.46
\textwidth]{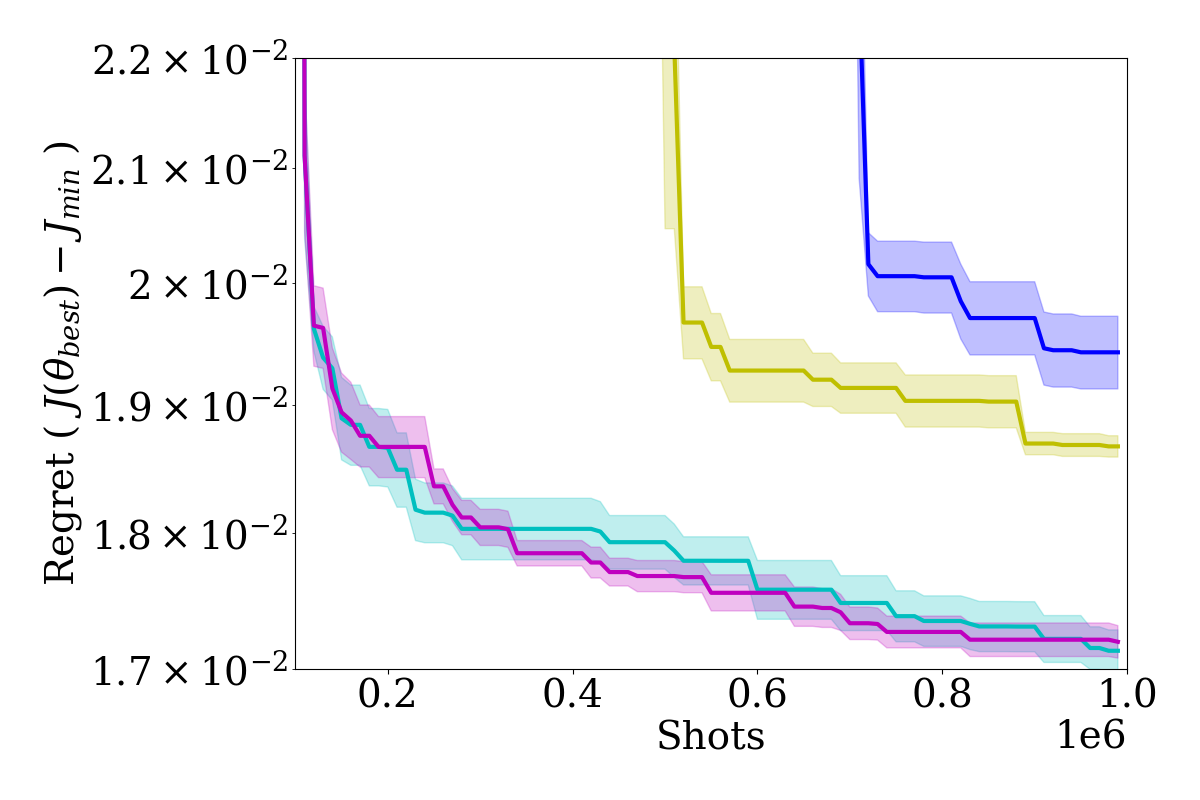}
     \caption{ Full ablation study results. Both the periodic kernel and the LSR independently improve the BO performance relative to a Vanilla strategy. The zoomed-in plot (right) shows that the combination of the LSR and periodic kernel provides the best performance, although it is a modest improvement over the LSR alone.  }
     \label{fig:full_ablation}
\end{figure}
Additionally, we present the results of the complete ablation in Fig. \ref{fig:full_ablation}, comparing the base strategy with a \matern kernel to one that uses a periodic kernel and two LSR strategies using \matern and periodic kernels. We note that the vanilla strategies can be considered a particular case of the low-shot residual where $r=1$. The vanilla periodic kernel (yellow) provides a clear advantage over the vanilla \matern kernel (blue), and both LSR strategies provide a marked advantage over the vanilla strategies. Although the difference between the two kernel options in the LSR is minuscule, the periodic kernel maintains slightly lower regret and lower variance across the runs.

\section{Conclusion}\label{sec:conclusion}
This work proposes two modifications to the standard Bayesian optimization implementation to improve shot-based efficiency when solving variational quantum algorithms. We show that a significant increase in performance can be achieved by encoding priors into the GP kernel function and surrogate model. The kernel prior endows the GP with knowledge of the parameter's 2$\pi$ periodicity, which we find helpful in the limited circuit observation regime. At the same time, the topological prior provides a better starting model by utilizing large quantities of low-shot circuit measurements. 

We remark on several directions for future work. First, although Bayesian optimization does not get stuck in local minima, it struggles to converge. The modifications discussed here will likely see similar performance gains by switching to local optimization methods to finish the optimization loop, as mentioned in \cite{Muller22}. Additionally, the consistent performance gains of LSR strategies with smaller $r$ values suggest that further reduction of the number of shots would improve performance; however, this could not be verified here due to computational issues in fitting the GP. To this end, further Gaussian process modification for large data sets would be worth exploring. Lastly, we note that these methods need further exploration and verification on more complex noise models, more challenging variational problems, and quantum hardware.

\section{Acknowledgement}
{\small{ We are grateful for support from NASA Ames Research Center. We acknowledge the funding of the NASA ARMD Transformational Tools and Technology (TTT) Project. 
F.S. and D.C. participated in the NASA/USRA Feynman
Quantum Academy internship program. 
D.B.N., F.S., and D.C. were supported by the NASA Academic Mission Services, Contract No. NNA16BD14C.
F.S. was partially supported by the NSF Graduate Research Fellowship. 
J.P. acknowledges funding from the National Science Foundation, Award No. 2237616.}}

{\small\bibsep=0pt
\bibliography{ref}
}
\end{document}